\documentclass[10pt]{article}

\usepackage{amsmath}
\usepackage{amssymb}

\usepackage{graphicx}

\usepackage{cite}

\usepackage{color} 

\usepackage{setspace} 
\usepackage[labelfont=bf,labelsep=period,justification=raggedright]{caption}

\makeatletter
\renewcommand{\@biblabel}[1]{\quad#1.}
\makeatother

\date{}

\begin{document}

% Title must be 150 characters or less
\begin{flushleft}
{\Large
\textbf{Markov Logic Networks in the Analysis of Genetic Data}
}
% Insert Author names, affiliations and corresponding author email.
\\
Nikita A. Sakhanenko,
David J. Galas$^{\ast}$
\\
Institute for Systems Biology, Seattle, WA, USA
\\
$\ast$ E-mail: dgalas@systemsbiology.org
\end{flushleft}

% Please keep the abstract between 250 and 300 words
\section*{Abstract}
Complex, non-additive genetic interactions are common and can be
critical in determining phenotypes. Genome-wide association studies
(GWAS) and similar statistical studies of linkage data, however,
assume additive models of gene interactions in looking for
associations between genotype and phenotype.  In general, these
statistical methods view the compound effects of multiple genes on a
phenotype as a sum of partial influences of each individual gene and
can often miss a substantial part of the heritable effect.  Such
methods do not make use of any biological knowledge about underlying
genotype-phenotype mechanisms. Modeling approaches from the Artificial
Intelligence field that incorporate deterministic knowledge into
models while performing statistical analysis can be applied to include
prior knowledge in genetic analysis.  We chose to use the most general
such approach, Markov Logic Networks (MLNs), that employs first-order
logic as a framework for combining deterministic knowledge with
statistical analysis.  Using simple, logistic regression-type MLNs we
have been able to replicate the results of traditional statistical
methods. Moreover, we show that even with quite simple models we are
able to go beyond finding independent markers linked to a phenotype by
using joint inference that avoids an independence assumption.  The
method is applied to genetic data on yeast sporulation, a phenotype
known to be governed by non-linear interactions between genes. In
addition to detecting all of the previously identified loci associated
with sporulation, our method is able to identify four additional loci
with small effects on sporulation. Since their effect on sporulation
is small, these four loci were not detected with standard statistical
methods that do not account for dependence between markers due to gene
interactions. We show how gene interactions can be detected using more
complex models, which in turn can be used as a general framework for
incorporating systems biology with genetics.  Such future work that
embodies systems knowledge in probabilistic models is proposed.

% Please keep the Author Summary between 150 and 200 words
% Use first person. PLoS ONE authors please skip this step. 
% Author Summary not valid for PLoS ONE submissions.   
\section*{Author Summary}

We  have  taken up  the  challenge of  devising  a  framework for  the
analysis  of  genetic data  that  is  fully  functional in  the  usual
statistical  correlation  analysis  used  in  genome-wide  association
studies,  but  also capable  of  incorporating  prior knowledge  about
biological systems  relevant to the genetic phenotypes.   We develop a
general genetic analysis approach that meets this challenge.  We adapt
an AI method  for learning models, called Markov  Logic Networks, that
is based on the fusion of Markov Random Fields with first order logic.
Our adaption  of the Markov  Logic Network method for  genetics allows
very complex  constraints and  a wide variety  of model classes  to be
imposed on probabilistic, statistical analysis.  We illustrate the use
of the method by analyzing  a data set based on sporulation efficiency
from yeast, in  which we demonstrate gene interactions  and identify a
number of new loci involved in determining the phenotype.

\section*{Introduction}

Genome-wide association  studies (GWAS) have allowed  the detection of
many  genetic  contributions  to  complex phenotypes  in  humans  (see
\emph{www.genome.gov}).  Studies  of biological networks  of different
kinds,   including   genetic   regulatory  networks,   protein-protein
interaction  networks and others,  have made  it clear,  however, that
gene interactions are abundant  and are therefore of likely importance
for   genetic    analysis~\cite{Manolio09}.    Complex,   non-additive
interactions  between  genetic  variations  are very  common  and  can
potentially     play     a     crucial     role     in     determining
phenotypes~\cite{Brem05,Drees05,Carter07,CarterDudley09}.    GWAS  and
similar statistical  methods such  as classical QTL  studies generally
assume additive models  of gene interaction that attempt  to capture a
compound effect of  multiple genes on a phenotype as  a sum of partial
influences of each individual gene~\cite{HirschhornDaly05,McCarthy08}.
These statistical  methods also  assume no biological  knowledge about
the underlying processes or phenotypes.  Since biological networks are
complex, and since variations are numerous, unconstrained searches for
associations between  genotype and phenotype  require large population
samples, and can succeed only in detecting a limited range of effects.
Without  imposing  any   constraints  based  on  biological  knowledge
searching for gene interactions is very challenging, particularly when
input  data  consist  of  different  data types  coming  from  various
sources.

The  major  question  that  motivated  this  work  is  ``\emph{Can  we
  constrain  traditional statistical  approaches  by using  biological
  knowledge to  define some known networks that  influence patterns in
  the  data, and  can such  approaches produce  more  complete genetic
  models?}''  For  example, we might  use the patterns present  in the
genotype data to  build more predictive models based  on both genotype
and  phenotype  data.   Note  that  the problem  of  using  biological
knowledge  to constrain  a  model of  genetic  interaction is  closely
connected to  the problem  of integrating various  types of data  in a
single  model.    In  this  article  we  employ   a  known  Artificial
Intelligence (AI) approach (Markov  Logic Networks) to reformulate the
problem of  defining and finding genetic  models in a  general way and
use  it to  facilitate  detection of  non-additive gene  interactions.
This  approach  allows  us  to  lay the  foundations  for  studies  of
essentially  any kind  of genetic  model, which  we demonstrate  for a
relatively simple model.

Markov Logic Networks (MLNs) is  one of the most general approaches to
statistical relational learning, a sub-field of machine learning, that
combines two kinds of modeling: probabilistic graphical models, namely
Markov Random Fields,  and first-order logic.  Probabilistic graphical
models,  first  proposed  by  Pearl~\cite{Pearl88},  offer  a  way  to
represent joint probability distributions  of sets of random variables
in  a   compact  fashion.    A  graphical  structure   describing  the
probabilistic  independence relationships in  these models  allows the
development  of numerous  algorithms  for learning  and inference  and
makes these models a good choice for handling uncertainty and noise in
data.  On the other hand, first-order logic allows us to represent and
perform  inferences over  complex, relational  domains.  Propositional
(Boolean) logic,  which biologists  are most familiar  with, describes
the truth state on the  level of specific instances, while first-order
logic allows us to make  assertions about the truth state of relations
between subsets  (classes) of instances.   Moreover, using first-order
logic we  can represent recursive and  potentially infinite structures
such as Markov chains where a temporal dependency of the current state
on  the state  at the  previous time  step can  be instantiated  to an
infinite  time series.   Thus, first  order logic  is a  very flexible
choice for  representing general knowledge, like that  we encounter in
biology.

MLNs merge  probabilistic graphical models and first-order  logic in a
framework  that  gains  the  benefits of  both  representations.  Most
importantly,  the logic component  of MLNs  provides an  interface for
adding biological  knowledge to a  model through a set  of first-order
constraints. At the same time, MLNs can be seen as a generalization of
probabilistic graphical  models since any  distribution represented by
the latter can  be represented by the former,  and this representation
is  more compact  due to  the  first-order logic  component. Even  so,
various learning and  inference algorithms for probabilistic graphical
models  are applicable  to MLNs  and are  thereby enhanced  with logic
inference.

One key  advantage of logic-based probabilistic  modeling methods, and
in particular  MLNs, is that  they allow us  to work easily  with data
that  are not  independent and  identically distributed  (not i.i.d.).
Many statistical  and machine learning  methods assume that  the input
data is i.i.d.,  a very strong, and usually  artificial, property that
most  biological  problems  do  not share.  For  instance,  biological
variables most often have a spatial or temporal structure, or can even
be  explicitly  described  in  a  relational  database  with  multiple
interacting    relations.    MLNs   thus    provide   a    means   for
non-i.i.d. learning  and joint inference  of a model. While  the input
data used  in GWAS and  in other genetic  studies are rich  in complex
statistical interdependencies between the data points, MLNs can easily
deal with any of these data structures.

There are  various modeling techniques that  employ both probabilistic
graphical              models              and             first-order
logic~\cite{Poole93,NgoHaddawy97,GlesnerKoller95,SatoKameya97,DeRaedt07,Friedman99,KerstingDeRaedt00,Pless06,RichardsonDomingos06}.
Many of  them impose different restrictions on  the underlying logical
representation in order  to be able to map  the logical knowledge base
to a  graphical model.  One common restriction  employed, for example,
in~\cite{SatoKameya97,DeRaedt07,KerstingDeRaedt00,Pless06}  is  to use
only \emph{clausal}  first-order formulas of  the form $b_1  \land b_2
\land  \ldots \land  b_n \Rightarrow  h$  that are  used to  represent
cause-effect  relationships.  The  majority  of the  methods, such  as
those                                                        introduced
in~\cite{NgoHaddawy97,GlesnerKoller95,KerstingDeRaedt00}, use Bayesian
networks,   directed   graphical    models,   as   the   probabilistic
representation.        However,       there       are      a       few
approaches~\cite{Pless06,RichardsonDomingos06} that instead use Markov
Random Fields, undirected graphical models, to perform inferences.

We  use Markov  Logic Networks~\cite{RichardsonDomingos06}  that merge
unrestricted  first-order logic with  Markov Random  Fields, and  as a
result  use  the   most  general  probabilistic  logic-based  modeling
approach.   In   this  paper  we  show  this   MLN-based  approach  to
understanding complex  systems and data  sets.  Similar to~\cite{Yi05}
that proposed a Bayesian approach that can be used to infer models for
QTL detection, our MLN-based approach is a model inference method that
goes  beyond just  hypothesis  testing.  Moreover,  in  this paper  we
describe how  we have adapted and  applied MLN to  genetic analysis so
that complex  biological knowledge can  be included in the  models. We
have applied the method to a relatively simple genetic system and data
set, the  analysis of  the genetics of  sporulation efficiency  in the
budding  yeast  \emph{Saccharomyces   cerevisiae}.   In  this  system,
recently   analyzed  by   Cohen  and   co-workers~\cite{Gerke09},  two
genetically  and phenotypically diverse  yeast strains,  whose genomes
were fully characterized, were crossed and the progeny studied for the
genetics of  the sporulation  phenotype.  This provided  a genetically
complex  phenotype with  a well-defined  genetic context  to  which to
apply our method.

% You may title this section "Methods" or "Models". 
% "Models" is not a valid title for PLoS ONE authors. However, PLoS ONE
% authors may use "Analysis" 
\section*{Methods}

\subsection*{Markov Random Fields}

Given a set of random  variables of the same type, $\mathbf{X}=\{X_i:1
\le   i  \le   N\}$  and   a   set  of   possible  values   (alphabet)
$\mathbf{A}=\{A_i:1 \le i  \le M\}$ so that any  variable can take any
value from $A_1$  to $A_M$ (it is  easy to extend this to  the case of
multiple  variable  types).  Consider  a  graph,  $G$, whose  vertices
represent   variables,  $\mathbf{X}$,   and   whose  edges   represent
probabilistic dependencies among the vertices such that a local Markov
property is met. The local Markov property for a random variable $X_i$
can  be formally  written  as $\Pr(X_i=A_j  \mid \mathbf{X}  \setminus \{X_i\})=\Pr(X_i=A_j \mid N(X_i))$ 
that states  that a  state of random  variable $X_i$  is conditionally
independent of  all other variables given  $X_i$'s neighbors $N(X_i)$,
$N(X_i)  \subseteq \mathbf{X}  \setminus  \{X_i\}$.  Let  $\mathbf{C}$
denote the  set of all  cliques in $G$,  where a clique is  a subgraph
that  contains  an  edge for  every  pair  of  its nodes  (a  complete
subgraph).  Consider  a configuration, $\gamma$,  of $\mathbf{X}$ that
assigns each  variable, $X_i$, a  value from $\mathbf{A}$.   We denote
the        space         of        all        configurations        as
$\mathbf{\Gamma}=\mathbf{A}^{\mathbf{X}}$.  A  restriction of $\gamma$
to the variables of a specific clique $C$ is denoted by $\gamma_C$.

A \emph{Markov  Random Field}  (MRF) is defined  on $\mathbf{X}$  by a
graph $G$  and a set of  potentials $\mathbf{V} =  \{ V_C(\gamma_C): C
\in \mathbf{C}, \gamma \in \mathbf{\Gamma} \}$ assigned to the cliques
of  the  graph. Using  cliques  allows  us  to explicitly  define  the
topology  of  models,  making  MRFs convenient  to  model  long-range,
higher-order   connections   between    variables.   We   encode   the
relationships between  the variables using the  clique potentials.  By
the  Hammersley-Clifford  theorem,  a joint  probability  distribution
represented by an MRF is given by the following Gibbs distribution
\begin{equation}
  \Pr(\gamma) = \frac{1}{Z} \prod_{C \in \mathbf{C}} \exp \left(-V_C(\gamma_C) \right),
  \label{eq1}
\end{equation}
where  the   so-called  partition  function  $Z   =  \sum_{\gamma  \in
  \mathbf{\Gamma}}  \prod_{C   \in  \mathbf{C}}  \exp(-V_C(\gamma_C))$
normalizes   the  probability   to  ensure   that   $\sum_{\gamma  \in
  \mathbf{\Gamma}} \Pr(\gamma) = 1$.

Without loss of generality we can represent a Markov Random Field as a
log-linear model~\cite{Pietra97}: 
\begin{equation}
  \Pr(\gamma)=\frac{1}{Z} \exp \left(\sum_i w_i f_i(\gamma) \right),
  \label{eq2}
\end{equation}
where  $f_i:  \mathbf{\Gamma}  \rightarrow \mathbb{R}$  are  functions
defining features of the MRF  and $w_i \in \mathbb{R}$ are the weights
of the MRF.   Usually, the features are indicators  of the presence or
absence of some attribute, and hence are binary.  For instance, we can
consider  a  feature  function that  is  $1$  when  some $X_i$  has  a
particular value and $0$  otherwise.  Using these types of indicators,
we can  make $M$  different features  for $X_i$ that  can take  on $M$
different   values.   Given  some   configuration  $\gamma_{\mathbf{X}
  \setminus \{X_i\}}$ of all  the variables $\mathbf{X}$ except $X_i$,
we can have  a different weight for this  configuration whenever $X_i$
has a  different value.   The weights for  these features  capture the
affinity of the  configuration $\gamma_{\mathbf{X} \setminus \{X_i\}}$
for each value of $X_i$. Note that the functions defining features can
overlap in arbitrary ways providing representational flexibility.

One simple mapping of a traditional  MRF to a log-linear MRF is to use
a  single feature  $f_i$ for  each configuration  $\gamma_C$  of every
clique $C$  with the weight $w_i  = - V_C(\gamma_C)$.   Even though in
this   representation  the   number   of  features   (the  number   of
configurations)  increases  exponentially   as  the  size  of  cliques
increases,  the Markov Logic  Networks described  in the  next section
attempt  to  reduce the  number  of  features  involved in  the  model
specification   by   using   logical   functions   of   the   cliques'
configurations.

\begin{figure}[!ht]
  \begin{center}
    \includegraphics[width=4in]{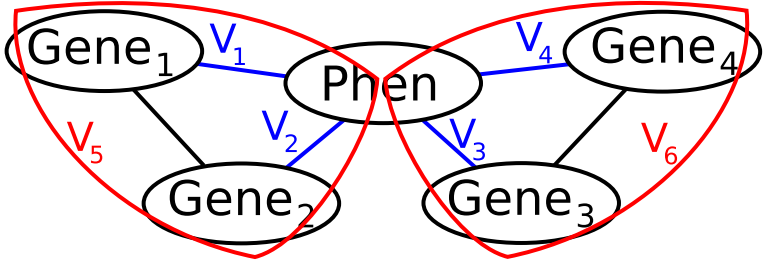}
  \end{center}
  \caption{\textbf{An  example  of a  Markov  Random  Field.} The  MRF
    represents  four  genes influencing  a  phenotype with  potentials
    $V_1,  \ldots, V_4$  (blue edges).   This model  restricts genetic
    interactions to  two pair-wise interactions  with potentials $V_5,
    V_6$ (red cliques).}
  \label{fig1}
\end{figure}

Given an  MRF, a general problem  is to find  a configuration $\gamma$
that  maximizes  the   probability  $\Pr(\gamma)$.   Since  the  space
$\mathbf{\Gamma}$ is  very large,  performing an exhaustive  search is
intractable.   For   many  applications,   there  are  two   kinds  of
information available:  prior knowledge about  the constraints imposed
on  the   simultaneous  configuration  of   connected  variables;  and
observations about  these variables for  a particular instance  of the
problem. The constraints constitute the model of the world and reflect
statistical dependencies  between values of the  neighbors captured in
an MRF.   For example, when modeling gene  association with phenotype,
the restrictions  on the likelihood of  configurations of co-expressed
genes  may be  cast  as an  MRF  with cliques  of size  2  and 3  (see
figure~\ref{fig1}). In the next  section, we give a biological example
involving  construction of  MRFs with  cliques of  size 3  and  4, and
provide more mathematical details.

\subsection*{Markov Logic Networks: Mapping First-Order Logic to Markov Random Fields}

Markov  Logic Networks  merge  Markov Random  Fields with  first-order
logic.  In  first-order logic (FOL)  we distinguish \textbf{constants}
and \textbf{variables}  that represent objects and their  classes in a
domain,  as  well as  \textbf{functions}  specifying mappings  between
subgroups of  objects, and \textbf{predicates}  representing relations
among objects or their  attributes.  We call a predicate \emph{ground}
if  all its variables  are assigned  specific values.   To illustrate,
consider  a   study  of  gene  interactions   through  the  phenotypic
comparison of  wild type strains,  single mutants, and  double mutants
such  as  the one  presented  in~\cite{Drees05}.   Consider  a set  of
constants representing genes, $\{ \mathtt{g1,g2} \}$, gene interaction
labels, $\{ \mathtt{A,B} \}$,  and difference between phenotype values
of  mutants,   $\{  \mathtt{0,1,2}  \}$.   We   define  the  following
predicates: $\mathtt{RelWS/2}$ (a  2-argument predicate which captures
a   relation   between   a   wild   type   and   a   single   mutant),
$\mathtt{RelWD/3}$  (a  relation between  a  wild  type  and a  double
mutant), $\mathtt{RelSS/3}$  (a relation between  two single mutants),
$\mathtt{RelSD/4}$ (a  relation between a  single mutant and  a double
mutant), $\mathtt{Int/3}$  (an interaction between  two genes).  Using
FOL we can define a knowledge base consisting of two formulas:
$$
\begin{array}{l}
  \displaystyle \mathtt{\forall x,y \in \{g1,g2\},\forall c \in \{0,1\},\forall v,u \in \{0,1,2\},
    Int(x,y,c) \Rightarrow (RelWS(x,v) \Leftrightarrow RelSD(y,x,y,u))}\\
  \displaystyle \mathtt{\forall x,y \in \{g1,g2\},\forall c \in \{0,1\},\forall v,u,w \in \{0,1,2\},
    RelWS(x,v) \land RelWS(y,u) \land RelWD(x,y,w) \Rightarrow Int(x,y,c).}   
\end{array}
$$
The first rule represents the  knowledge that depending on the type of
interaction  between   two  genes,  there  is   a  dependency  between
$\mathtt{RelWS(x,v)}$  and  $\mathtt{RelSD(y,x,y,u)}$ relations.   The
second   rule   captures   the   knowledge   that   three   relations,
$\mathtt{RelWS(x,v)}$,            $\mathtt{RelWS(y,u)}$,           and
$\mathtt{RelWD(x,y,w)}$,   together  determine   the   type  of   gene
interaction.

Note that  first-order formulas define  relations between (potentially
infinite) groups of objects or  their attributes.  Formulas in FOL can
be   seen  as   relational  templates   for  constructing   models  in
propositional  logic.    Therefore,  FOL  offers  a   compact  way  of
representing  and  aggregating  relational  data.   For  example,  two
first-order  formulas above  can  be replaced  with 288  propositional
formulas since variables $\mathtt{x,y,c}$  can be assigned 2 different
values  and  variables  $\mathtt{u,v,w}$  can be  assigned  3  values.
Moreover, using representational power of FOL, we can specify infinite
structures such  as temporal relations,  e.g., $\mathtt{Expression(e1,
  t1)  \land NextTimeStep(t1,  t2)  \Rightarrow Expression(e2,  t2)}$,
that can give rise to a theoretically infinite number of propositions.

The principal limitation  of any strictly formal logic  system is that
it  is   not  suitable  for  real  applications   where  data  contain
uncertainty and  noise.  For  example, the formulas  specified earlier
hold for the real data most of  the time, but not always.  If there is
at least one data point where a formula does not hold, then the entire
model is disregarded  as being false. The two  allowed states, true or
false, is equivalent  to allowing only probability values  $1$ or $0$.
Markov Logic Networks, however,  relax this constraint by allowing the
model  with unsatisfied formula  with a  lesser probability  than one.
The model with  the smallest number of unsatisfied  formulas will then
be the most probable.

Markov Logic Networks (MLNs) extend  FOL by assigning a weight to each
formula indicating its probabilistic strength.  An MLN is a collection
of first-order formulas $F_i$  with associated weights $w_i$. For each
variable of a Markov Logic Network  there is a finite set of constants
representing  the domain  of  the variable.   A  Markov Logic  Network
together with its corresponding constants is mapped to a Markov Random
Field as follows.

Given a set of all predicates on an MLN, every ground predicate of the
MLN corresponds to one random  variable of a Markov Random Field whose
value  is $1$  if  the ground  predicate  is true  and $0$  otherwise.
Similarly, every ground formula of $F_i$ corresponds to one feature of
the log-linear  Markov Random Field whose  value is $1$  if the ground
formula is true  and $0$ otherwise.  The weight of  the feature in the
Markov Random  Field is the  weight $w_i$ associated with  the formula
$F_i$ in the Markov Logic Network.

From the original definitions (\ref{eq1}) and (\ref{eq2}) and the fact
that  features  of  false  ground  formulas  are  equal  to  $0$,  the
probability distribution  represented by a  \emph{ground} Markov Logic
Network is given by
\begin{equation}
  \Pr(\gamma)=\frac{1}{Z} \exp \left(\sum_i w_i n_i(\gamma)\right) 
  = \frac{1}{Z} \prod_i \exp \left(-V_i(\gamma_i)n_i(\gamma)\right),
  \label{eq3}
\end{equation}
where $n_i(\gamma)$ is a number of true ground formulas of the formula
$F_i$ in the state $\gamma$  (which directly corresponds to our data),
$\gamma_i$  is  the configuration  (state)  of  the ground  predicates
appearing in $F_i$. $V_i$ is a potential function assigned to a clique
which  corresponds  to  $F_i$,  and  $\exp(-V_i(\gamma_i))=\exp(w_i)$.
Note that this probability distribution would change if we changed the
original  set of  constants.  Thus,  one  can view  MLNs as  templates
specifying classes  of Markov Random  Fields, just like  FOL templates
specifying propositional formulas.

\begin{figure}[!ht]
  \begin{center}
    \includegraphics[width=4in]{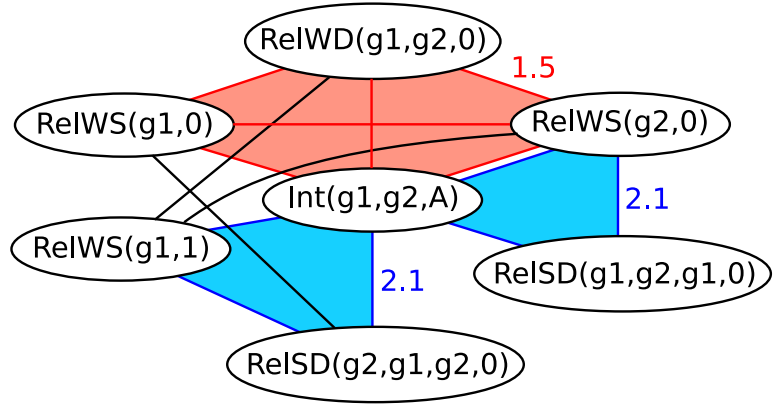}
  \end{center}
  \caption{  \textbf{An example  of a  subnetwork of  a  Markov Random
      Field unfolded  from a Markov Logic Network  program.} Each node
    of the  MRF corresponds  to a ground  predicate (a  predicate with
    variables  substituted with  constants). The  nodes of  all ground
    predicates that appear  in a single formula form  a clique such as
    the  one  highlighted  with   red.  The  blue  triangular  cliques
    correspond to  the first formula of  the MLN and  are assigned the
    weight of the formula  (2.1). The larger rectangular cliques, such
    as the  one colored red, correspond  to the second  formula of the
    MLN with the weight 1.5.}
  \label{fig2}
\end{figure}

Figure~\ref{fig2}  illustrates  a portion  of  a  Markov Random  Field
corresponding to the  ground MLN. We assume a set  of constants and an
MLN specified  by the knowledge base  from the example  above, where a
weight is assigned to each formula.

\subsection*{An Example: Application to Yeast Sporulation Dataset}

We  applied  our  method  to   a  data  set  generated  by  Cohen  and
co-workers~\cite{Gerke09}.  They generated  and characterized a set of
374 progeny of a cross between two yeast strains that differ widely in
their efficiency of  sporulation (a wine and an  oak strain). For each
of the progeny the sporulation  efficiency was measured and assigned a
normalized real value  between 0 and 1.  To  generate a discrete value
set  we then  binned and  mapped the  sporulation efficiencies  into 5
integer  values.   Each yeast  progeny  strain  was  genotyped at  225
markers that are uniformly  distributed along the genome.  Each marker
takes on one of two possible values indicating whether it derived from
the oak or wine parent genotype.

Using Markov  Logic Networks,  we first model  the effect of  a single
marker  on  the phenotype,  i.e.,  sporulation  efficiency.  Define  a
logistic-regression type model with the following set of formulas:
\begin{equation}
  \mathtt{\forall strain \in \{1, \ldots ,374\}, \hspace{5pt} G(strain,m,g) 
    \Rightarrow E(strain,v), \hspace{5pt} w_{m,g,v}},
  \label{eq4}
\end{equation}
for every  marker $\mathtt{m}$ under  consideration (at this  point we
consider one  marker in  a model), genotype  value $\mathtt{g}  \in \{
\mathtt{A},\mathtt{B}\}$,   and   phenotype   value  $\mathtt{v}   \in
\{\mathtt{1},  \ldots,  \mathtt{5}\}$.    This  Markov  Logic  Network
contains  two predicates,  $\mathtt{G}$  and $\mathtt{E}$.   Predicate
$\mathtt{G}$  denotes markers' genotype  values across  yeast crosses,
e. g., $\mathtt{G(strain,M1,B)}$ captures  all yeast crosses for which
the genotype  of a  marker $\mathtt{M1}$ is  $\mathtt{B}$.  Similarly,
predicate $\mathtt{E}$ denotes  the phenotype (sporulation efficiency)
across  yeast crosses,  for instance,  $\mathtt{E(strain,1)}$ captures
all yeast  strains for  which the level  of sporulation  efficiency is
$\mathtt{1}$.   The  Markov  Logic  Network  (\ref{eq4})  contains  10
formulas, 1 marker of interest  times 2 possible genotype values times
5 possible  phenotype values.  Each formula represents  a pattern that
holds  true  across  all  yeast  crosses (indicated  by  the  variable
$\mathtt{strain}$)  with the  same strength  (indicated by  the weight
$\mathtt{w_{m,g,v}}$).      In     other     words,     the     weight
$\mathtt{w_{m,g,v}}$  represents  the  fitness  of  the  corresponding
formula across all strains.

Instantiations  of  the  predicate  $\mathtt{G}$ represent  a  set  of
predictor   variables,  whereas   instantiations   of  the   predicate
$\mathtt{E}$ represent  a set of target  variables (\ref{eq4}).  There
are 748 ground predicates of $\mathtt{G}$ (assuming we handle only one
marker in a  model) and 1870 ground predicates  of $\mathtt{E}$.  Each
ground predicate corresponds to a random variable in the corresponding
Markov  Random Field  (see  the previous  section  for more  details).
Since  our  MLN  contains  10  formulas and  there  are  374  possible
instantiations for  each formula, the  corresponding log-linear Markov
Random  Field contains 3740  features, one  for each  instantiation of
every formula.

\subsection*{Learning the Weights of MLNs}

Each  data point  in the  original dataset  corresponds to  one ground
predicate (either  $\mathtt{E}$ or $\mathtt{G}$ in  our example).  For
example,  the   information  that  a   genotype  value  of   a  marker
$\mathtt{M71}$  in a  strain $\mathtt{S13}$  is equal  to $\mathtt{A}$
corresponds to a ground predicate $\mathtt{G(S13,M71,A)}$.  Therefore,
the original  dataset can be  represented with a collection  of ground
predicates that logically  hold, which in turn is  described as a data
vector $\mathbf{d} =  \langle d_1, \ldots, d_N \rangle$,  where $N$ is
the  number  of  all  possible  ground predicates  ($N=2618$  in  this
example).  An  element $d_i$  of the vector  $\mathbf{d}$ is  equal to
$1$, if the  $i$th ground predicate (assuming some  order) is true and
thus is included in our collection, and $0$ otherwise.  Note that this
vector   representation  is  possible   under  a   \emph{closed  world
  assumption} stating  that all ground predicates that  are not listed
in our collection are assumed to be false.

In order  to carry out training of  a Markov Logic Network  we can use
standard Newtonian methods  for likelihood maximization.  The learning
proceeds by iteratively improving weights  of the model.  At the $j$th
step,     given     weights     $\mathbf{w}^{(j)}$,     we     compute
$\nabla_{\mathbf{w}^{(j)}}  L(\mathbf{w}^{(j)})$, the gradient  of the
likelihood,  which  is  our   objective  function  that  we  maximize.
Consequently, we improve the weights by moving in the direction of the
positive  gradient, $\mathbf{w}^{(j+1)}  =  \mathbf{w}^{(j)} +  \alpha
\nabla_{\mathbf{w}^{(j)}} L(\mathbf{w}^{(j)})$,  where $\alpha$ is the
step size.

Recall that the likelihood is given by
\begin{equation}
  L(\mathbf{w} \mid \gamma) 
  = \Pr(\gamma \mid \mathbf{w})
  = \frac{1}{Z} \exp \left(\sum_i w_i n_i(\gamma)\right)
  = \frac{\exp \left(\sum_i w_i n_i(\gamma)\right)}
         {\sum_{\gamma' \in \mathbf{\Gamma}} \exp \left(\sum_i w_i n_i(\gamma')\right)},
  \label{eq5}
\end{equation}
where $\gamma$ is a state (also  called a configuration) of the set of
random variables $\mathbf{X}$ and  $\mathbf{\Gamma}$ is a space of all
possible states.  Therefore, the log-likelihood is
\begin{equation}
  \log L(\mathbf{w} \mid \gamma)
  = \log \Pr(\gamma \mid \mathbf{w})
  = \sum_i w_i n_i(\gamma) 
  - \log \left[ \sum_{\gamma' \in \mathbf{\Gamma}} \exp\left(\sum_i w_i n_i(\gamma')\right) \right].
  \label{eq6}
\end{equation}

Now derive the gradient with respect to the network weights,
\begin{equation}
  \begin{array}{l}
    \displaystyle \frac{\partial}{\partial w_j} \log L(\mathbf{w} \mid \gamma) 
    = n_j(\gamma) - \frac{1}{\sum_{\gamma' \in \mathbf{\Gamma}} \left[ Z \Pr(\gamma') \right]} 
    \frac{\partial}{\partial w_j} \sum_{\gamma' \in \mathbf{\Gamma}} \exp \left(\sum_i w_i n_i(\gamma')\right)\\
    \displaystyle = n_j(\gamma) - \frac{1}{\sum_{\gamma' \in \mathbf{\Gamma}} \left[ Z \Pr(\gamma') \right]} 
    \sum_{\gamma' \in \mathbf{\Gamma}} \left[ n_j(\gamma')\exp \left(\sum_i w_i n_i(\gamma')\right) \right]\\
    \displaystyle = n_j(\gamma) 
    - \sum_{\gamma' \in \mathbf{\Gamma}} \left[ n_i(\gamma')L(\mathbf{w} \mid \gamma') \right].
  \end{array}
  \label{eq7}
\end{equation}

Note that the sum is computed over \emph{all possible} variable states
$\gamma'$.   The above  expression shows  that each  component  of the
gradient  is a  difference  between  the number  of  true instances  a
corresponding  formula $F_j$ (the  number of  true ground  formulas of
$F_j$) and the expected number of true instances of $F_j$ according to
the  current model.  However,  computation of  both components  of the
difference is intractably large.

Since the  exact number  of true ground  formulas cannot  be tractably
computed   from  data   \cite{RichardsonDomingos06},  the   number  is
approximated  by sampling the  instances of  the formula  and checking
their truth values according to the data.

On  the other hand,  it is  also intractable  to compute  the expected
number  of  true  ground   formulas  as  well  as  the  log-likelihood
$L(\mathbf{w} \mid \gamma')$.  The  former involves inference over the
model, whereas the later  requires computing the partition function $Z
=   \sum_{\gamma'   \in   \mathbf{\Gamma}}   \exp   \left(\sum_i   w_i
n_i(\gamma')       \right)$.        One       solution,       proposed
in~\cite{RichardsonDomingos06},      is      to      maximize      the
\emph{pseudo-likelihood}
\begin{equation}
  \hat{L}(\mathbf{w} \mid \gamma) = \prod_{j=1}^N \Pr(\gamma_j \mid \gamma_{MB_j};\mathbf{w}),
  \label{eq8}
\end{equation}
where $\gamma_j$ is  a restriction of the state  $\gamma$ to the $j$th
ground predicate  and $\gamma_{MB_j}$ is a restriction  of $\gamma$ to
what is  called a  Markov blanket of  the $j$th ground  predicate (the
state of the Markov blanket according to our data).  We elected to use
this  approach.  Similar  to  the original  definition  of the  Markov
blanket  in  the  context  of Bayesian  networks  \cite{Pearl88},  the
\emph{Markov blanket} of  a ground predicate is a  set of other ground
predicates that are  present in some ground formula.   Using the yeast
sporulation example, the set of ground predicates $\{ \mathtt{ \forall
  m,  \forall  g  \mid  G(S1,m,g)  }  \}$ is  the  Markov  blanket  of
$\mathtt{E(S1,1)}$   due   to    the   knowledge   base   (\ref{eq4}).
Maximization  of pseudo-likelihood  is computationally  more efficient
than maximization  of likelihood, since it does  not involve inference
over the model, and thus does not require marginalization over a large
number   of   variables.   Currently,   we   use  the   limited-memory
Broyden-Fletcher-Goldfarb-Shanno    (L-BFGS)   algorithm    from   the
\emph{Alchemy}  implementation of  MLNs \cite{Kok09}  to  optimize the
pseudo-likelihood.

\subsection*{Using MLNs for Querying}

After learning  is complete,  we have a  trained Markov  Logic Network
that can  be used for various  types of inference.   In particular, we
can  answer such  queries as  ``\emph{what is  the probability  that a
  ground predicate $Q$  is true given that every  predicate from a set
  (a conjunction)  of ground  predicates $Ev=\{E_1, \ldots,  E_m\}$ is
  true?}'' The ground  predicate $Q$ is called a  \emph{query} and the
set $Ev$ is called an \emph{evidence}. Answering this query is similar
to computing the  probability $\Pr(Q \mid E_1 \land  \ldots \land E_m,
MLN)$.  Using the product rule for probabilities we get
\begin{equation}
  \begin{array}{l}
    \displaystyle \Pr(Q \mid E_1 \land \ldots \land E_m, MLN) 
    = \frac{\Pr(Q \land E_1 \land \ldots \land E_m \mid MLN)}{\Pr(E_1 \land \ldots \land E_m \mid MLN)}\\
    \displaystyle = \frac{\sum_{\gamma \in \mathbf{\Gamma}_Q \cap \mathbf{\Gamma}_E} \Pr(\gamma \mid MLN)}
                         {\sum_{\gamma \in \mathbf{\Gamma}_E} \Pr(\gamma \mid MLN)},
  \end{array}
  \label{eq9}
\end{equation}
where   $\mathbf{\Gamma}_P$   is  the   set   of  \emph{all   possible
  configurations}  where   a  ground   predicate  $P$  is   true,  and
$\mathbf{\Gamma}_E   =    \mathbf{\Gamma}_{E_1}   \cap   \ldots   \cap
\mathbf{\Gamma}_{E_m}$.

Computing  the   above  probabilistic  query  for   majority  of  real
application problems,  which include the  computational problems posed
by  the complexity  experienced  in systems  biology, is  intractable.
Therefore, we need  to approximate $\Pr(Q \mid E_1  \land \ldots \land
E_m, MLN)$, which can be done using various sampling-based algorithms.

Markov   Logic   Networks    adopt   a   \emph{knowledge-based   model
  construction} approach consisting of  two steps: 1. constructing the
smallest Markov Random Field from  the original MLN that is sufficient
for  computing the  probability of  the  query, and  2. inferring  the
Markov Random Field using traditional approaches.  One of the commonly
used  inference algorithms  is Gibbs  sampling where  at each  step we
sample a ground predicate $X_j$ given its Markov blanket.  In order to
define the probability of the node $X_j$ being in the state $\gamma_j$
given the state of its Markov blanket we use the earlier notation.

Given  a $j$th  ground predicate  $X_j$, all  the  formulas containing
$X_j$ are denoted  by $\mathbf{F}_j$. We denote the  Markov blanket of
$X_j$ as  $MB_j$ and a restriction  of $\gamma$ to  the Markov blanket
(the  state of the  Markov blanket)  as $\gamma_{MB_j}$.  Similarly, a
restriction   of  the   state  $\gamma$   to  $X_j$   is   denoted  as
$\gamma_j$. Recall  that each  formula $F$ of  a Markov  Logic Network
corresponds to a feature of a Markov Random Field, where the feature's
value is  the truth value $f$  of the formula $F$  depending on states
$\gamma_1, \ldots,  \gamma_k$ of  the ground predicates  $X_1, \ldots,
X_k$ constituting the formula  and denoted by $f=F|_{\gamma_1, \ldots,
  \gamma_k}$. Note that $F|_{\gamma_1,  \ldots, \gamma_k}$ can also be
shown as  $F|_{\gamma_1, \gamma_{MB_j}}$.  Using this notation  we can
express  the  probability  of  the  node  $X_j$ to  be  in  the  state
$\gamma_j$ when its Markov blanket is in the state $\gamma_{MB_j}$ as
\begin{equation}
  \Pr(\gamma_j \mid \gamma_{MB_j}) 
  = \frac{\exp \left( \sum_{F_i \in \mathbf{F}_j} w_i F_i|_{\gamma_j, \gamma_{MB_j}} \right)}
  {\exp \left( \sum_{t=0}^1 \sum_{F_i \in \mathbf{F}_j} w_i F_i|_{X_j=t, \gamma_{MB_j}} \right)}.
  \label{eq10}
\end{equation}

For  the Gibbs  sampling,  we let  a  Markov chain  converge and  then
estimate the probability  of a conjunction of ground  predicates to be
true  by   counting  the  fraction  of  samples   from  the  estimated
distribution in which all the ground predicates hold. The Markov chain
is  run  multiple  times  in  order  to  handle  situations  when  the
distribution has  multiple local maxima  so that the Markov  chain can
avoid  being  trapped  on  one  of  the peaks.   One  of  the  current
implementations,   called  \emph{Alchemy}~\cite{Kok09},   attempts  to
reduce  the burn-in  time of  the Gibbs  sampler by  applying  a local
search  algorithm  for  the  weighted satisfiability  problem,  called
MaxWalkSat~\cite{Selman93}.

\subsection*{Components of the Computational Method}

\begin{figure}[!ht]
  \begin{center}
    \includegraphics[width=0.99\textwidth]{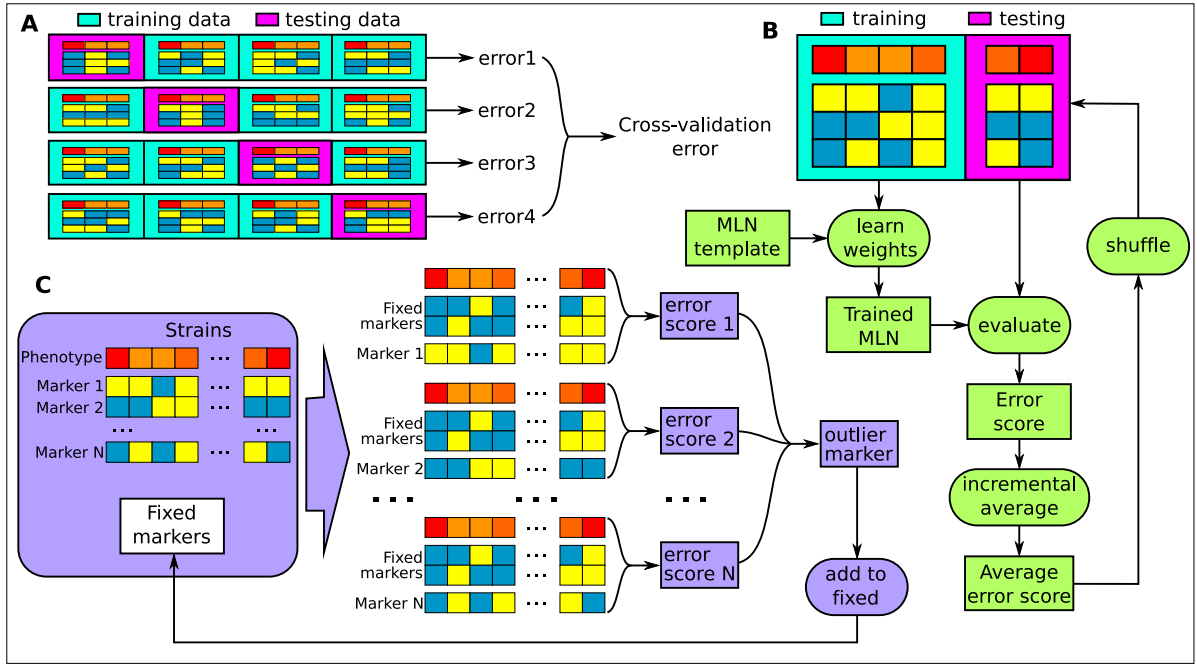}
  \end{center}
  \caption{  \textbf{Components of  the  computational method.}   This
    figure illustrates three  major computational components.  We used
    cross-validation to estimate the goodness of fit of a model. Panel
    \textbf{A}   depicts  4-fold   cross-validation.    The  data   is
    partitioned  into four  sets  and  at each  iteration  a model  is
    trained on three sets and  then tested on the fourth set resulting
    in four prediction error scores that are then averaged for a total
    cross-validation  error. Panel \textbf{B}  shows details  of model
    training and testing. Using  training data we estimate the weights
    of each formula of an  MLN template.  The trained MLN is evaluated
    on testing data resulting in an error score.  We shuffle the order
    in the training data and the  order of the testing data and repeat
    the training and testing.   The error scores after each evaluation
    are averaged to produce the  average error score of the MLN. Panel
    \textbf{C}  shows   how  the  components   illustrated  in  panels
    \textbf{A}  and \textbf{B}  are combined  to search  for  the most
    informative markers.  At the  $N$th iteration of the search, given
    a data  set and set of  $N-1$ fixed markers,  the method traverses
    the set of all markers  and evaluates models constructed using the
    fixed  markers and  a  selected  marker. Note  that  at the  $N$th
    iteration, $N$-marker models consider the fixed markers ($N-1$) as
    possible  interactors with the  next marker  we found.  The method
    then selects  the outlier model, adds the  corresponding marker to
    the  set  of fixed  markers,  and  repeats  the traversal  of  the
    markers. The method stops when no outliers are found.}
  \label{fig3}
\end{figure}

The  general  overview  of   the  computational  method  is  given  in
figure~\ref{fig3}.  At the first step, the method traverses the set of
all markers and  assigns an error score to  each marker indicating its
predictive  power.   An  error   score  of  a  marker  corresponds  to
performance  of an  MLN (\ref{eq4})  based  on this  single marker  (a
random variable  $\mathtt{m}$ of (\ref{eq4})  that essentially chooses
the location of markers  to use in the model has only  one value – the
location of this single  marker).  The algorithm selects markers whose
error scores are considerably lower  than the average: we selected the
outliers that are 3 standard deviations below the mean.

The current version  of the method greedily selects  a marker with the
lowest  score and  appends it  to a  list of  fixed markers  (which is
initially empty). The algorithm then repeats the search for an outlier
with  the lowest score,  but this  time a  marker's score  is computed
using an MLN  (\ref{eq4}) that estimates a joint  effect of the marker
under consideration  together with all of the  currently fixed markers
on a  phenotype (the variable  $\mathtt{m}$ takes on the  locations of
all  fixed  markers  and  the  marker under  consideration).  At  each
iteration the algorithms expands the list of fixed markers and rescans
all of the remaining markers for outliers. The method stops as soon as
no  more  outliers are  detected  and  returns  the fixed  markers  as
potential loci associated with the phenotype.

Our scanning for predictive genetic markers can be seen as an instance
of the  \emph{variable selection} problem. We  use cross-validation to
compare probabilistic models  and select the one with  the best fit to
the  data  and the  smallest  number  of  informative markers.   Using
cross-validation we  assess how a model generalizes  to an independent
dataset,  addressing the  model overfitting  problem  and facilitating
unbiased   outlier   detection.    Particularly,   we   use   $K$-fold
cross-validation         which        is         illustrated        in
figure~\ref{fig3}(\textbf{A}).  The data set is arbitrarily split into
$K$  folds, $\mathbf{D}_1, \ldots,  \mathbf{D}_K$, and  $K$ iterations
are  performed.   The  $i$th  iteration  of  cross-validation  selects
$\bigcup_{j   \ne  i}   \mathbf{D}_j$  as   a  training   dataset  and
$\mathbf{D}_i$ as a testing dataset.  The model is then trained on the
training dataset  and the performance  of the model is  assessed using
the testing dataset  resulting in a prediction error.   The average of
prediction errors from $K$  steps, called a cross-validation error, is
used as a score of the model.  In case of yeast sporulation efficiency
dataset introduced earlier, we  used 11-fold cross-validation (since a
population  of 374  yeast strains  can be  evenly partitioned  into 11
subsets  with  34  strains   in  each).  The  results  were  generally
insensitive to the cross-validation parameters.

Recall that  we distinguish two  types of variables,  the \emph{target
  variables}  whose  values  are  predicted, and  the  \emph{predictor
  variables} whose values are used to predict the values of the target
variables.  In  the example above,  sporulation efficiency of  a yeast
strain is  a target variable,  whereas genotype markers  are predictor
variables.  Note  that in  some cases we  can treat variables  as both
targets  and  predictors (e.g.   gene  expression  in eQTL  datasets).
During the evaluation phase in the $i$th iteration of cross-validation
we consider the  testing dataset $\mathbf{D}_i$. Using knowledge-based
model construction  approach, we build  a Markov Random Field  that is
small yet  sufficient to infer the  values of all  target variables in
the set $\mathbf{D}_i$. The target variables are inferred based on the
values  of the  predictor variables  from $\mathbf{D}_i$  (see section
``Using MLNs for Querying'').

The model  prediction of a  target variable $X$  that can take  on any
value  from   $\{x_1,x_2,x_3\}$  can   be  represented  as   a  vector
$\hat{\mathbf{v}}=\langle  p_1,  p_2,   p_3  \rangle$,  where  $p_j  =
\Pr(X=x_j \mid \mathbf{D}_i, \Theta_{MRF})$  is the probability of $X$
to take on a value $x_j$ given the testing data $\mathbf{D}_i$ and the
Markov Random Field with parameters $\Theta_{MRF}$. On the other hand,
the   actual  value   of  $X$   (provided  in   the   testing  dataset
$\mathbf{D}_i$)  can be  represented as  a  vector $\mathbf{v}=\langle
v_1, v_2, v_3 \rangle$, where $\forall j \ne k, v_j = 0$ and $v_k = 1$
iff  $X=x_k$  in $\mathbf{D}_i$.   Then  the  prediction error  should
measure the  difference between the  prediction $\hat{\mathbf{v}}$ and
the  true   value  $\mathbf{v}$.   We  used  the   Euclidean  distance
$d(\hat{\mathbf{v}},\mathbf{v})$  to  compute  the  prediction  error.
This approach might make  the comparison to other approaches difficult
since the error can be a value that is not bounded by $0$ and $1$, but
by $0$ and $\sqrt{M}$,  where $M$ is the size of the  domain of $X$ (3
in our example).  Further computation is required to obtain values for
model   accuracy,   to    explain   variance,   and   other   standard
characteristics. On  the other  hand, Euclidean distance  is certainly
sufficient for comparing predictions of different models.

Due to the  approximate nature of learning and  inference in MLNs (see
sections  ``Learning  the  Weights  of  MLNs'' and  ``Using  MLNs  for
Querying''), two  structurally identical  models, trained on  two data
sets  that differ  only  in the  order  of the  samples, can  generate
predictions with  slight differences.  This is due  to the fundamental
path-dependency    of   learning    and   inference    algorithms   in
knowledge-based  model  construction.    For  example,  the  order  of
training data  affects the order in  which the Markov  Random Field is
built,  which in  turn affects  the way  the approximate  reasoning is
performed over the field.  Path-dependency introduces artificial noise
into predictions and considerably reduces our ability to distinguish a
signal with  a small magnitude (such  as a possible minor  effect of a
genetic locus on a phenotype) from a background.

\begin{figure}[!ht]
  \begin{center}
    \includegraphics[width=3in]{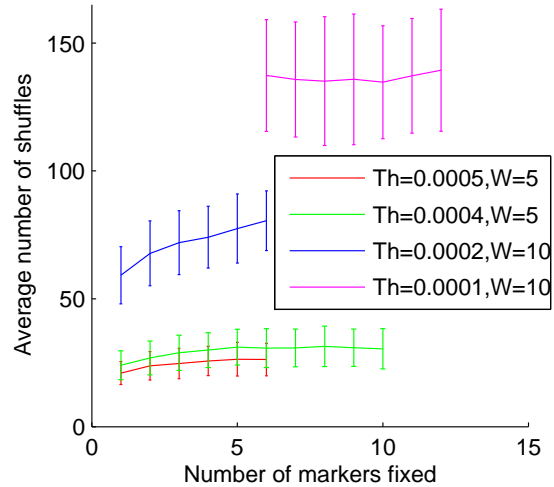}
  \end{center}
  \caption{ \textbf{The  amount of shuffling depends  on the threshold
      and the size of the  window.} Note that the number of iterations
    until convergence  of the  algorithm increases when  the threshold
    decreases or the window  size increases.  Note also that selecting
    tight stopping parameters tends to allow the algorithm to identify
    more informative markers.}
  \label{fig4}
\end{figure}

In  order to  reduce the  effect of  path-dependency on  overall model
prediction we  shuffle the  input data set  and average  the resulting
predictions.  We  employed an iterative approach,  based on shuffling,
for  denoising.   At  each   iteration  the  model  is  retrained  and
reevaluated on newly  shuffled data and the running  mean of the model
prediction  is   computed  (see  figure~\ref{fig3}(\textbf{B})).   The
method incrementally  computes the prediction  average until achieving
convergence, namely  until the difference between  the running average
at  the  two consecutive  iterations  is  smaller  than $Th$  for  $W$
consecutive steps.  The parameters $Th$ and $W$ are directly connected
with  the total  amount of  shuffling and  re-estimation  performed as
illustrated in figure~\ref{fig4}.

In order  to perform rigorous denoising,  we select a  lower value for
the threshold  $Th$ ($0.0001$)  and a larger  window size  $W$ ($10$).
The shuffling-based  denoising procedure is applied  at each iteration
of  the cross-validation.   Averaging  of the  predictions after  data
shuffling reduces the amount  of artificial noise enabling the overall
method for  detection of  genetic loci to  distinguish markers  with a
smaller  effect   on  the   phenotype  (the  algorithm   detects  more
informative markers as illustrated in figure~\ref{fig4}).

There are many different strategies to search for the most informative
subset of genetic  markers. In this section we  used a greedy approach
in  order  to  illustrate  the general  MLN-based  modeling  framework
presented in  this paper. In the  next section we  show that MLN-based
modeling  that  accounts  for  dependencies  between  markers  through
joint-inference allows  us to  find interesting biological  results by
using a greedy search method. In  order to be confident that the fixed
markers are meaningful, we manually selected markers at each iteration
of the search from the set of outliers and arrived at a similar set of
candidate loci (within the same local region).

% Results and Discussion can be combined.
\section*{Results}

The  analyses  presented  in  this  paper are  based  on  the  dataset
from~\cite{Gerke09} containing both phenotype (sporulation efficiency)
and   genotype  information   of  yeast   strains  derived   from  one
intercross. The  results are obtained  using our method  that searches
for the  largest set  of genetic markers  with the  strongest compound
effect  on  the  phenotype.   All  the  detailed  information  on  the
computational  components of our  method is  presented in  the Methods
section.

In their  paper Gerke et al.~\cite{Gerke09} identified  5 markers that
have  an effect on  sporulation efficiency  including 2  markers whose
effect  seems to  be very  small.   Moreover, Gerke  et al.   provided
evidence  for  non-linear  interactions  between 3  major  loci.   The
presence  of  confirmed markers  with  various  effect and  non-linear
interactions make the dataset  from~\cite{Gerke09} an ideal choice for
testing our computational method.

\begin{figure}[!ht]
  \begin{center}
    \includegraphics[width=0.99\textwidth]{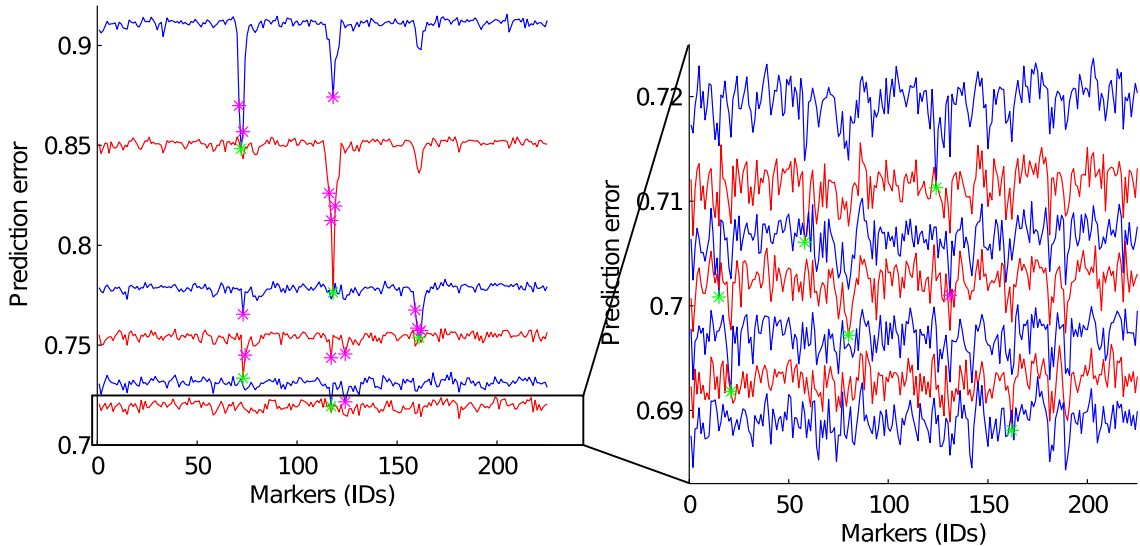}
  \end{center}
  \caption{  \textbf{Prediction  of  sporulation  efficiency  using  a
      subset  of markers.}   This figure  shows the  execution  of the
    algorithm  plotted as  error scores  vs  markers for  a number  of
    models. The  top line in the  left plot shows error  scores of the
    models based only  on a single marker plotted  on the $X$-axis.  A
    green star  indicates the outlier  marker with the  smallest error
    and  purple  stars  depict  other outliers  (markers  whose  error
    differs from the mean error by 3 standard deviations).  The second
    line shows the error scores of the models based on a corresponding
    marker  together with  the previously  selected  marker (indicated
    with the green star).  All  the following lines are interpreted in
    the similar  way. The  left plot shows  five markers with  a large
    effect.  The  rest of the  identified markers (six markers  with a
    small effect) are illustrated in the right plot.}
  \label{fig5}
\end{figure}

Our  method allows us  to define  and to  use essentially  any genetic
model. First we used a simple regression-type model that mimics simple
statisical  approaches, like  GWAS.  At  the first  stage,  the method
compares the  markers according to their  individual predictive power.
The amount of the effect of  a marker on the phenotype is estimated by
computing  a prediction  error  score from  a  regression model  based
solely on  this marker.  The top line  on the  left panel in  figure 5
illustrates the error  scores of all markers ordered  by location ($X$
axis). In figure~\ref{fig5} we  observe three loci (around markers 71,
117,  and 160) with  the strongest  effect on  sporulation efficiency,
which were identified in~\cite{Gerke09}.

At  the next  stage, our  method adds  the marker  with  the strongest
effect (marker  71) to  all the following  models.  This allows  us to
compare all other markers according to their predictive power \emph{in
  conjunction}  with the  fixed marker  71. This  time  the prediction
error score of a marker indicates how well this marker \emph{together}
with the marker 71 predicts the sporulation efficiency.  The score is,
thus, computed  from the  regression model based  on two  markers: the
marker 71 and  a second marker. This is  an important distinction from
traditional GWAS  where the searches for  multiple influential markers
are performed  independently from each  other. In our  approach, using
MLNs and in particular joint-inference, the compound effect of markers
is estimated allowing us to see possible interactions between markers.

The method continues the iterations  and selects 11 markers before the
error  no  longer improves  sufficiently  and  the computation  stops.
Among the selected markers, 5  are the same loci previously identified
in~\cite{Gerke09} (markers 71,  117, 160, 123, 79), 3  are the markers
next to the loci with the strongest effect (markers 72, 116, 161), and
3 are the new markers that  have not been reported before (markers 57,
14,  20). In addition,  the method  identifies another  marker (marker
130) as  a candidate  for a  locus that has  an effect  on sporulation
efficiency,  although this  marker was  not selected  due to  its weak
predictive power.   Notice that even  with a relatively  simple model,
such  as logistic  regression,  and a  quite  stringent criterion  for
outliers (3 standard deviations from the mean, a $p$-value $0.003$ for
a normal distribution) we are  able to exceed the number of identified
candidate  loci.   We argue  that  our  method  is more  efficient  at
discovering  markers with a  very low  individual effect  on phenotype
that  have non-trivial  interactions with  other sporulation-affecting
loci due to the use of joint-inference of MLNs.

There are several distinct properties of our method that are important
to note. First, although the method selects the neighboring markers of
the three  strongest loci, it  does not select a  neighbor immediately
after the original loci has  been identified, because there are better
markers to be found. For example, after selecting the first marker 71,
the method finds markers 117 and 160, and only then selects marker 72,
which is  the neighbor  of 71. The  method selects the  next strongest
marker at each  stage that maximally increases the  compound effect of
selected markers. Second, our method does not find markers that do not
add sufficient predictive power. The criterion for outliers determines
when the  method stops  and determines the  confidence that  the added
markers have a real effect on the phenotype.

For each new marker (57, 14,  20, 130) we examined all genes that were
nearby (different  actual distances were  used).  For example  for the
marker  14 we  considered all  genes  located between  markers 13  and
15. Table~\ref{tab1}  shows genes located near  these newly identified
markers that are involved in either meiosis or in sporulation.
\begin{table}[!ht]
  \caption{
    \bf{Candidate genes near the new informative markers.}}
  \begin{tabular}{|c|p{40pt}|p{30pt}|p{45pt}|c|p{220pt}|}
    \hline
    Marker & Euclidean Error & Coord. & Candidate Genes & GO & Description\\
    \hline
    57 & 0.7061 & 6, 103743 & YFL039C, ACT1 & S & Actin, structural protein involved in cell polarization, endocytosis, and other cytoskeletal functions\\
    & & & YFL037W, TUB2 & M & Beta-tubulin; associates with alpha-tubulin (Tub1p and Tub3p) to form tubulin dimer, which polymerizes to form microtubules\\
    & & & YFL033C, RIM15 & M & Glucose-repressible protein kinase involved in signal transduction during cell proliferation in response to nutrients, specifically the establishment of stationary phase; identified as a regulator of IME2; substrate of Pho80p-Pho85p kinase\\
    & & & YFL029C, CAK1 & M & Cyclin-dependent kinase-activating kinase required for passage through the cell cycle, phosphorylates and activates Cdc28p\\
    & & & YFL009W, CDC4 & M & F-box protein required for G1/S and G2/M transition, associates with Skp1p and Cdc53p to form a complex, SCFCdc4, which acts as ubiquitin-protein ligase directing ubiquitination of the phosphorylated CDK inhibitor Sic1p\\
    & & & YFL005W, SEC4 & S & Secretory vesicle-associated Rab GTPase essential for exocytosis\\
    \hline
    14 & 0.7009 & 2, 656824 & YBR180W, DTR1 & S & Putative dityrosine transporter, required for spore wall synthesis; expressed during sporulation; member of the major facilitator superfamily (DHA1 family) of multidrug resistance transporters\\
    & & & YBR186W, PCH2 & M & Nucleolar component of the pachytene checkpoint, which prevents chromosome segregation when recombination and chromosome synapsis are defective; also represses meiotic interhomolog recombination in the rDNA\\
    \hline
    130 & 0.7010 & 11, 447373 & YKR029C, SET3 & M & Defining member of the SET3 histone deacetylase complex which is a meiosis-specific repressor of sporulation genes; necessary for efficient transcription by RNAPII\\
    & & & YKR031C, SPO14 & S & Phospholipase D, catalyzes the hydrolysis of phosphatidylcholine, producing choline and phosphatidic acid; involved in Sec14p-independent secretion; required for meiosis and spore formation; differently regulated in secretion and meiosis\\
    \hline
    20 & 0.6972 & 3, 188782 & YCR033W, SNT1 & M & Subunit of the Set3C deacetylase complex that interacts directly with the Set3C subunit, Sif2p; putative DNA-binding protein\\
    \hline
  \end{tabular}
  \begin{flushleft}
    The list of  genes located near the new  markers identified by the
    MLN-based method. The table shows only the genes that are involved
    in sporulation or meiosis.  Specific information for the genes can
    be found at \emph{www.yeastgenome.org}.
  \end{flushleft}
  \label{tab1}
\end{table}

The  simple  logistic  regression-type  model  that was  used  can  be
summarized  using   the  first-order  formula  $\mathtt{G(strain,m,g)}
\Rightarrow  \mathtt{E(strain,v)}$,  which captures  the  effect of  a
subset  of markers  on phenotype.  In order  to  investigate gene-gene
interactions we used a pair-wise  model which can be summarized by the
formula   $\mathtt{G(strain,m1,g1)}   \land   \mathtt{G(strain,m2,g2)}
\Rightarrow \mathtt{E(strain,v)}$.   The pair-wise model  subsumes the
simple   regression   model,    since   whenever   $\mathtt{m1}$   and
$\mathtt{m2}$ are identical,  the pair-wise MLN is mapped  to the same
set of cliques  as those from the simple  MLN.  However, the pair-wise
model defines the  dependencies between two loci and  a phenotype that
are  mapped to  an additional  set of  3-node cliques.   The pair-wise
model allows  us \emph{explicitly} to  account for the  pair-wise gene
interactions. When  using the pair-wise model,  the joint-inference is
performed over an MLN  where possible interactions between two markers
are specified with first-order formulas.

The  assumption inherent  in  genome-wide analyses  is  that a  simple
additive effect can  be observed when applying the  pair-wise model to
loci that do not interact: the compound effect is essentially a sum of
the  individual effects  of each  locus. On  the other  hand,  for two
interacting  markers, the  pair-wise model  is expected  to  predict a
larger-than-additive  compound  effect.   Since  the  pair-wise  model
incorporates possible interactions, the prediction error of this model
should be  smaller than the error of  a simple model by  a factor that
corresponds to  how much the interaction information  helps to improve
the prediction.
\begin{figure}[!ht]
  \begin{center}
    \includegraphics[width=3in]{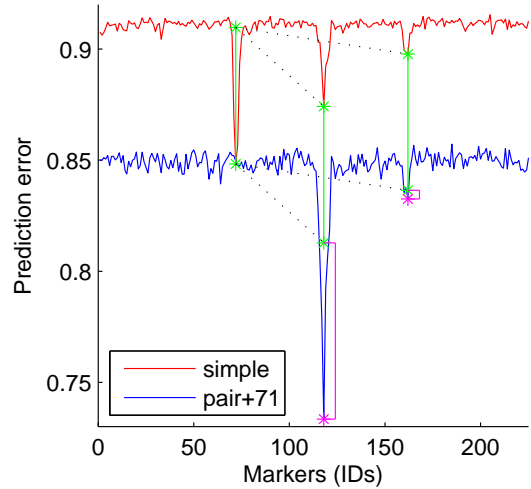}
  \end{center}
  \caption{ \textbf{Investigating the  71-117 loci interaction.}  This
    figure  compares a  standard  genome-wide scan  made  by a  simple
    regression model  based on a single  marker (red line)  and a scan
    made by  a pair-wise model based  on two markers, one  of which is
    preset to 71  (blue line).  The green lines  represent the size of
    the leftmost red peak  corresponding to the difference $d$ between
    the baseline  prediction error of  the simple model and  the error
    $error_S(71)$. Pink bars represent how much the difference between
    $error_S(117)$  and $error_{PW}(71,117)$ is  larger than  $d$. The
    large size of the leftmost pink bar indicates a strong interaction
    between markers 71 and 117.}
  \label{fig6}
\end{figure}

By  using  the  pair-wise  model,  we  investigated  the  presence  of
interactions  between markers 71,  117, 160  which correspond  to loci
with the  strongest effect on  sporulation efficiency.  We  denote the
prediction error of  a simple regression model based  on markers $M_1,
\ldots,  M_n$ as  $error_S(M_1,  \ldots,  M_n)$, and  the  error of  a
pair-wise   model   based   on   markers   $M_1,   \ldots,   M_n$   as
$error_{PW}(M_1,   \ldots,  M_n)$.   Figure~\ref{fig6}   compares  the
prediction errors of  the simple regression model based  on one marker
(red line) and the errors of  the pair-wise model based on two markers
one of  which is  preset to  71 (blue line).   Note that  the baseline
prediction error of the pair-wise  model is the same as $error_S(71)$,
which  means that  on average  the choice  of a  second marker  in the
pair-wise model does not affect the prediction.  There are, however, 2
markers that  visibly improve the  prediction, namely markers  117 and
160. Note  that the  prediction error $error_{PW}(71,160)$  (the right
blue peak)  is lower than  $error_S(160)$ (the rightmost red  peak) by
only  a value  roughly equal  to  the difference  between the  average
prediction errors of simple and  pair-wise models (this value is equal
to  the  size  of  the  leftmost  red peak).   The  reduction  of  the
prediction  error, when  combining markers  71 and  160,  is additive,
suggesting that there is no  interaction between these two markers. On
the other hand,  if we look at the effect of  combining markers 71 and
117, we  can see that  the prediction improvement using  the pair-wise
model  based on  both markers  (the  size of  the left  blue peak)  is
considerably more  than just a  sum of prediction improvements  of two
simple models  independently (the leftmost and the  middle red peaks).
The  non-additive improvement  suggests that  there is  an interaction
between the markers 71 and 117.
\begin{figure}[!ht]
  \begin{center}
    \includegraphics[width=3in]{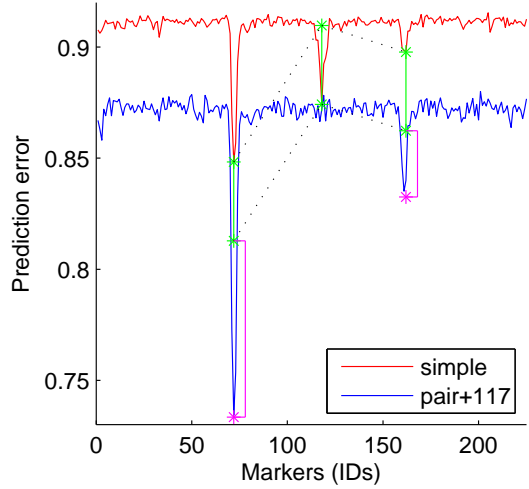}
  \end{center}
  \caption{   \textbf{Investigation  of   117-71   and  117-160   loci
      interactions.} This figure  compares a standard genome-wide scan
    using a  single-marker model  and a scan  using a  pair-wise model
    based on  two markers,  one of  which is preset  to 117.   See the
    caption  of figure~\ref{fig6}  for  more details.  Both pink  bars
    indicate the presence of non-additive interactions between markers
    117 and 71 and markers 117 and 160.}
  \label{fig7}
\end{figure}
\begin{figure}[!ht]
  \begin{center}
    \includegraphics[width=3in]{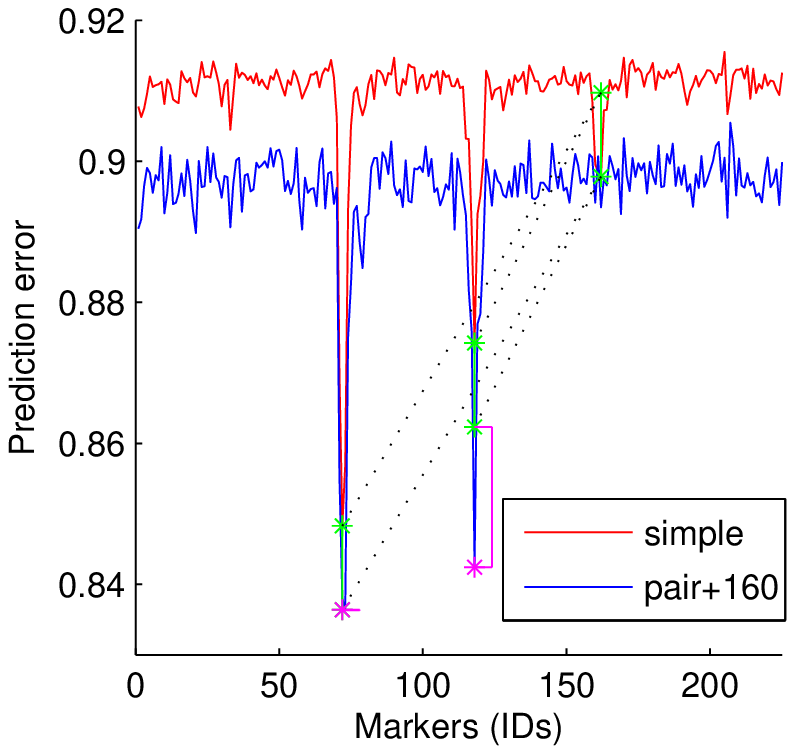}
  \end{center}
  \caption{ \textbf{Investigating  an interaction between  markers 117
      and 160.} This figure compares a standard genome-wide scan using
    a single-marker model and a  scan using a pair-wise model based on
    two markers,  one of which  is preset to  160. See the  caption of
    figure~\ref{fig6} for more details. Note $error_{PW}(71,160)$ that
    is   almost  the   same  as   $error_{PW}(117,160)$   even  though
    $error_S(71)$ is considerably  lower than $error_S(117)$. The tall
    pink bar on the right  side of the figure indicates a non-additive
    interaction between  markers 117 and  160. On the other  hand, the
    pink bar on the left  is almost non-existent indicating absence of
    an interaction between markers 71 and 160.}
  \label{fig8}
\end{figure}

Figures~\ref{fig7}  and \ref{fig8}  show  a similar  analysis to  that
illustrated   in  figure~\ref{fig6}   performed  on   the   other  two
markers.  The   analysis  shown  in   figure~\ref{fig7}  confirms  the
interactions between  markers 71  and 117 and,  additionally, suggests
that   there  is  an   interaction  between   markers  117   and  160.
Figure~\ref{fig8} confirms the interaction between 117 and 160 and the
absence  of the  interaction  between 71  and  160. One  can see  from
figure~\ref{fig8}   that    the   leftmost   blue    peak   indicating
$error_{PW}(71,160)$ is a sum of $error_S(71)$ and $error_S(160)$ (the
pink bar next to the left  pink star is extremely short). On the other
hand,  the rightmost  blue  peak is  a lot  more  than just  a sum  of
individual    errors   (the    pink   bar    is   tall).    In   fact,
$error_{PW}(117,160)$ is almost the same as $error_{PW}(71,160)$.

\begin{figure}[!ht]
  \begin{center}
    \includegraphics[width=0.99\textwidth]{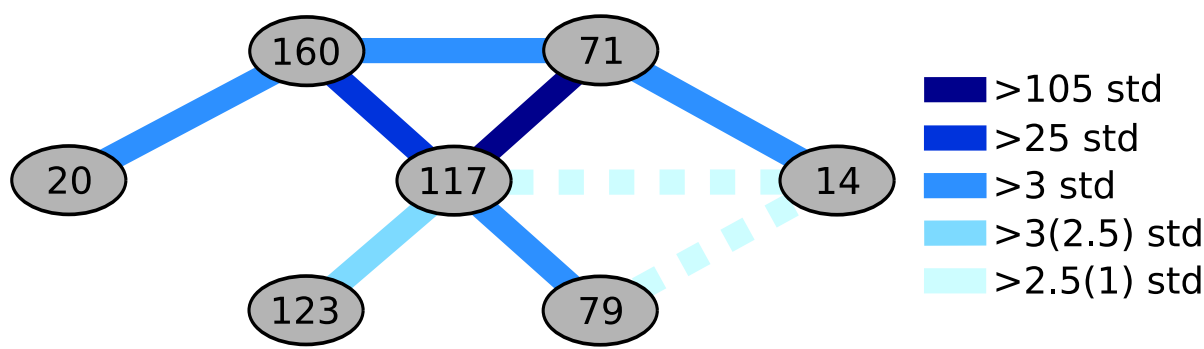}
  \end{center}
  \caption{ \textbf{Estimated network of gene-gene interactions.} This
    figure  shows a  network  of estimated  interactions between  loci
    based on genetic data and  sporulation phenotype. The color of the
    links corresponds to the  level of confidence of interactions. The
    three  darkest   colors  are  associated  with   the  most  likely
    interactions. The  fourth color is associated  with an interaction
    123-117 that  scored high in  the first experiment (more  than $3$
    standard deviations away  from the mean), but lower  in the second
    experiment ($2.5$  standard deviations), although  still above the
    confidence  level  ($2.32$  standard  deviations).   Two  possible
    interactions  depicted with  the  dashed lines  with the  lightest
    color scored  above the confidence  level in the  first experiment
    ($2.54$ standard  deviations), but  below the confidence  level in
    the second experiment ($1$ standard deviation). Loci 160, 71, 117,
    123,  79 were  previously identified  in~\cite{Gerke09}. Moreover,
    the  genes of  the  three major  loci  were detected:  \emph{IME1}
    (locus 117),  \emph{RME1} (locus 71), and  \emph{RSF1} (locus 160)
    \cite{Gerke09}.  The  two   strongest  interactions,  160-117  and
    117-71, were also identified in~\cite{Gerke09}.}
  \label{fig9}
\end{figure}

The   two   predicted   interactions,   71-117   and   117-160,   were
experimentally  identified in~\cite{Gerke09}.   The strength  of these
interactions is significant enough to immediately stand out during the
analysis in  figure~\ref{fig6}. We next  applied this analysis  to the
set of all  nine identified loci (71, 117, 160, 123,  57, 14, 130, 79,
20) in order  to quantify possible interactions between  every pair of
markers. For  each two  markers $A$ and  $B$ from  the set of  loci we
compute the prediction errors of a simple model based solely on either
$A$ or $B$, denoted as  $error_S(A)$ and $error_S(B)$. We also compute
the prediction error  of a pair-wise model based on  both $A$ and $B$,
denoted  as $error_{PW}(A,B)$.  Consequently  the size  of a  possible
interaction  between $A$ and  $B$, denoted  as $i(A,B)$,  is estimated
using the following expression:
\begin{equation}
  \begin{array}{l}
    \displaystyle i(A,B) = d(A,B) - d(A) - d(B)\\
    \displaystyle = (median - error_{PW}(A,B)) - (median - error_S(A)) - (median - error_S(B))\\
    \displaystyle = error_S(A) + error_S(B) - error_{PW}(A,B) - median.
  \end{array}
  \label{eq11}
\end{equation}
Here  $median$ is a  baseline of  prediction error  of a  simple model
based  on  a  single  marker.    We  averaged  the  errors  over  $10$
independently computed  iterations.  We  next determined how  high the
value  $i(A,B)$   should  be  in  order  to   confidently  predict  an
interaction between  markers $A$ and  $B$.  We selected $36$  pairs of
randomly  selected  markers  that  were  not  from  the  set  of  nine
informative markers and computed $i(A,B)$  for each pair.  Since we do
not expect  any interactions between  random, non-informative markers,
their $i(A,B)$ values  are used to estimate a  confidence interval for
no interaction.  We computed a  mean and standard deviation of the set
of $i(A,B)$  values corresponding to the randomly  chosen markers.  It
is estimated that a value  of $2.54$ standard deviations away from the
mean  completely covers  the set  of  $i(A,B)$ values  for all  random
markers,  and we  therefore  argue  there is  a  strong likelihood  of
interaction between markers $A$ and $B$ whenever $i(A,B)$ is more than
$3$ standard deviations away from the mean.  Whenever $i(A,B)$ is less
than $3$ but more than  $2.54$ standard deviations away from the mean,
we  argue  there  is  a  probable interaction  between  $A$  and  $B$.
Estimated interactions  between identified  loci are illustrated  as a
network of marker interactions in figure~\ref{fig9} where the color of
each  link represents  the level  of confidence  of  the corresponding
interaction.  We  repeated the estimation of  interactions by randomly
selecting another  $36$ pairs of markers and  computing the confidence
intervals for the new  set ($2.32$ standard deviations).  The probable
interactions identified  from the  first experiment were  confirmed in
the  second  experiment.   Two  possible interactions,  however,  were
identified in  one experiment  but not the  other and are  depicted in
figure~\ref{fig9} with dashed  links.  This is a result  of a slightly
shifted mean of the second set of $36$ random marker pairs relative to
the first set, and the  marginal size of the effect.  Two interactions
with very  large $i(A,B)$ values, 71-117 and  117-160, were previously
identified   in~\cite{Gerke09}.   We   also   found  several   smaller
interactions  illustrated  in  figure~\ref{fig9}  that have  not  been
identified  before.   Note  that  since  we  measure  absolute  values
$i(A,B)$, it  is not a  surprise that interactions 71-117  and 117-160
are so large,  since the corresponding loci (71, 117,  160) are by far
the  strongest.   Locus  117,  which  is  involved  in  the  strongest
interactions, corresponds to  the gene \emph{IME1}~\cite{Gerke09}, the
master  regulator   of  meiosis  (\emph{www.yeastgenome.com}).   Since
\emph{IME1}  is a  very  important sporulation  gene,  it is  entirely
reasonable  that  this gene  is  central  to  our interaction  network
(figure~\ref{fig9}).

\section*{Discussion}

The  method presented  in this  paper provides  a framework  for using
virtually any genetic model in a genome-wide study because of the high
representational  power of  MLNs.  This  power stems  from the  use of
general,     first-order    logic    conjoined     to    probabilistic
reasoning. Moreover, the use  of knowledge-based model approaches that
build models  based on  both data and  a \emph{relevant} set  of first
order formulas~\cite{Wellman92}  allows us to  efficiently incorporate
prior biological  knowledge into a  genetic study.  The  generality of
MLNs  allows   greater  representational  power   than  most  modeling
approaches.   The  general  approach  can  be  viewed  as  a  seamless
unification  of statistical  analysis, model  learning  and hypothesis
testing.

As opposed to standard genome-wide approaches to genetics which assume
additivity, the aim of the method  described in this article is not to
return values  corresponding to the strength of  individual effects of
each  marker.   Our method  aims  at  discovering  the loci  that  are
involved in  determining the phenotype. The method  computes the error
scores for each  marker in the context of  the others representing the
strength of each marker's effect in combination with other markers. It
will be valuable  in future to derive a  scoring technique for markers
that can be  used directly to compare with  results of the traditional
approaches. In general,  our approach provides a way  of searching for
the best model  predicting the phenotype from the  genetic loci. Since
the   model   and   the  corresponding   joint-inference   methodology
incorporate the relations between the  model variables, we are able to
begin  a  quantitative exploration  of  possible interactions  between
genetic loci.

Our method  shows promise  in that it  can accommodate  complex models
with  internal  relationships  among  the  variables  specified.   The
development of a succinct and  clear language and grammar based on FOL
for  the description of  (probabilistic) biological  systems knowledge
will  be critical  for the  widespread application  of this  method to
genetic  analyses.    Achieving  this  goal  will   also  represent  a
significant step toward the fundamental integration of systems biology
and   the   analysis  and   modeling   of   networks  with   genetics.
Additionally,  the development of  the biological  language describing
useful biological  constraints can alleviate  the computational burden
associated with model inference. MLN-based methods, such as ours, that
perform  both logic and  probabilistic inferences  are computationally
expensive.   While increases  in computing  power steadily  reduce the
magnitude of  this problem,  there are other  approaches that  will be
necessary.  Given such a  focused biological language, we could tailor
the learning  and inference algorithms  specifically to our  needs and
thus  reduce  the  overall  computational complexity  of  the  method.
Another future direction will  be to find fruitful connections between
a  previously  developed  information  theory-based approach  to  gene
interactions~\cite{Carter09} with this  AI-derived approach. There are
clear  several other  applications of  this approach  in the  field of
biology.  It is  clear that there are, for  example, many similarities
between  the   problem  discussed  here  and  the   problems  of  data
integration.

Biomarker  identification, particularly for  panels of  biomarkers, is
another  important  problem  that  involves many  challenges  of  data
integration and that can  benefit from our MLN-based approach. Similar
to GWAS  or QTL  mapping, where  we search for  genetic loci  that are
linked with a phenotype of  interest, in biomarker detection we search
for proteins or  miRNAs that are associated with a  disease state in a
set of patient data.  Just as in genetics, we can represent biomarkers
as  a   network  because  there  are   various  underlying  biological
mechanisms that  govern the development  of a disease. Often  the most
informative markers from a medical  point of view have weak signals by
themselves.  MLNs can allow  us to incorporate partial knowledge about
the   underlying    biological   processes   to    account   for   the
inter-dependencies making the  detection of the informative biomarkers
more effective.  It is clear  that the approach described here has the
potential to  integrate the biomarker  problem with human  genetics, a
key problem in the future development of personalized medicine.

% Do NOT remove this, even if you are not including acknowledgments
\section*{Acknowledgments}

The authors are grateful to  Aimee Dudley, Barak Cohen and Greg Carter
for  stimulating   discussions  and  also  Barak   Cohen  for  sharing
experimental data.  This work  was supported by  a grant from  the NSF
FIBR  program to DG,  and the  University of  Luxembourg-Institute for
Systems Biology  Program. We also  thank Dudley, Tim  Galitski, Andrey
Rzhetsky and especially Greg Carter for comments on the manuscript.

%\section*{References}
% The bibtex filename
\bibliography{MLN-paper}

\begin{thebibliography}{10}
\providecommand{\url}[1]{\texttt{#1}}
\providecommand{\urlprefix}{URL }
\expandafter\ifx\csname urlstyle\endcsname\relax
  \providecommand{\doi}[1]{doi:\discretionary{}{}{}#1}\else
  \providecommand{\doi}{doi:\discretionary{}{}{}\begingroup
  \urlstyle{rm}\Url}\fi
\providecommand{\bibAnnoteFile}[1]{%
  \IfFileExists{#1}{\begin{quotation}\noindent\textsc{Key:} #1\\
  \textsc{Annotation:}\ \input{#1}\end{quotation}}{}}
\providecommand{\bibAnnote}[2]{%
  \begin{quotation}\noindent\textsc{Key:} #1\\
  \textsc{Annotation:}\ #2\end{quotation}}
\providecommand{\eprint}[2][]{\url{#2}}

\bibitem{Manolio09}
Manolio TA, Collins FS, Cox NJ, Goldstein DB, Hindorff LA, et~al. (2009)
  {Finding the missing heritability of complex diseases}.
\newblock Nature 461: 747--753.
\bibAnnoteFile{Manolio09}

\bibitem{Brem05}
Brem RB, Storey JD, Whittle J, Kruglyak L (2005) {Genetic interactions between
  polymorphisms that affect gene expression in yeast}.
\newblock Nature 436: 701--703.
\bibAnnoteFile{Brem05}

\bibitem{Drees05}
Drees BL, Thorsson V, Carter GW, Rives AW, Raymond MZ, et~al. (2005)
  {Derivation of genetic interaction networks from quantitative phenotype
  data}.
\newblock Genome Biology 6: R38.
\bibAnnoteFile{Drees05}

\bibitem{Carter07}
Carter GW, Prinz S, Neou C, Shelby JP, Marzolf B, et~al. (2007) {Prediction of
  phenotype and gene expression for combinations of mutations}.
\newblock Molecular Systems Biology 3: 96.
\bibAnnoteFile{Carter07}

\bibitem{CarterDudley09}
Carter GW, Dudley AM (2009) {Systems Genetics and Complex Traits}.
\newblock Encyclopedia of Complexity and Systems Science : 9105--9124.
\bibAnnoteFile{CarterDudley09}

\bibitem{HirschhornDaly05}
Hirschhorn JN, Daly MJ (2005) {Genome-wide association studies for common
  diseases and complex traits}.
\newblock Nature Reviews Genetics 6: 95--108.
\bibAnnoteFile{HirschhornDaly05}

\bibitem{McCarthy08}
McCarthy MI, Abecasis GR, Cardon LR, Goldstein DB, Little J, et~al. (2008)
  {Genome-wide association studies for complex traits: consensus, uncertainty
  and challenges}.
\newblock Nature Reviews Genetics 9: 356--369.
\bibAnnoteFile{McCarthy08}

\bibitem{Pearl88}
Pearl J (1988) {Probabilistic Reasoning in Intelligent Systems: Networks of
  Plausible Inference}.
\newblock San Francisco, CA: Morgan Kaufmann.
\bibAnnoteFile{Pearl88}

\bibitem{Poole93}
Poole D (1993) {Logic Programming, Abduction and Probability: a top-down
  anytime algorithm for estimating prior and posterior probabilities}.
\newblock New Generation Computing 11: 377--400.
\bibAnnoteFile{Poole93}

\bibitem{NgoHaddawy97}
Ngo L, Haddawy P (1997) {Answering queries from context-sensitive probabilistic
  knowledge bases}.
\newblock Theoretical Computer Science 171: 147--177.
\bibAnnoteFile{NgoHaddawy97}

\bibitem{GlesnerKoller95}
Glesner S, Koller D (1995) {Constructing Flexible Dynamic Belief Networks from
  First-Order Probabilistic Knowledge Bases}.
\newblock In: Proc. of the European Conference on Symbolic and Quantitative
  Approaches to Reasoning and Uncertainty. pp. 217--226.
\bibAnnoteFile{GlesnerKoller95}

\bibitem{SatoKameya97}
Sato T, Kameya Y (1997) {PRISM: a language for symbolic-statistical modeling}.
\newblock In: Proc. of the 15th Intl. Joint Conf. on AI (IJCAI). pp.
  1330--1335.
\bibAnnoteFile{SatoKameya97}

\bibitem{DeRaedt07}
DeRaedt L, Kimmig A, Toivonen H (2007) {ProbLog: a probabilistic Prolog and its
  application in link discovery}.
\newblock In: Proc. of the 20th Intl. Joint Conf. on AI (IJCAI). pp.
  2468--2473.
\bibAnnoteFile{DeRaedt07}

\bibitem{Friedman99}
Friedman N, Getoor L, Koller D, Pfeffer A (1999) {Learning Probabilistic
  Relational Models}.
\newblock In: Proc. of 16th Intl. Joint Conf. on AI (IJCAI). pp. 1300--1307.
\bibAnnoteFile{Friedman99}

\bibitem{KerstingDeRaedt00}
Kersting K, DeRaedt L (2000) {Bayesian Logic Programs}.
\newblock In: Cussens J, Frisch A, editors, Proc. of 10th Int. Conf. on ILP.
  pp. 138--155.
\bibAnnoteFile{KerstingDeRaedt00}

\bibitem{Pless06}
Pless DJ, Chakrabarti C, Rammohan R, Luger GF (2006) {The Design and Testing of
  a First-Order Stochastic Modeling Language}.
\newblock International Journal on Artificial Intelligence Tools 15: 979--1005.
\bibAnnoteFile{Pless06}

\bibitem{RichardsonDomingos06}
Richardson M, Domingos P (2006) {Markov Logic Networks}.
\newblock Machine Learning 62: 107--136.
\bibAnnoteFile{RichardsonDomingos06}

\bibitem{Yi05}
Yi N, Yandell BS, Churchill GA, Allison DB, Eisen EJ, et~al. (2005) {Bayesian
  Model Selection for Genome-Wide Epistatic Quantitative Trait Loci Analysis}.
\newblock Genetics 170: 1333--1344.
\bibAnnoteFile{Yi05}

\bibitem{Gerke09}
Gerke J, Lorenz K, Cohen B (2009) {Genetic Interactions between Transcription
  Factors Cause Natural Variation in Yeast}.
\newblock Science 323: 498--501.
\bibAnnoteFile{Gerke09}

\bibitem{Pietra97}
Pietra SD, Pietra VD, Lafferty J (1997) {Inducing Features of Random Fields}.
\newblock IEEE Trans of Pattern Analysis and Machine Intelligence 19: 380--393.
\bibAnnoteFile{Pietra97}

\bibitem{Kok09}
Kok S, Sumner M, Richardson M, Singla P, Poon H, et~al. (2007) {The Alchemy
  system for statistical relational AI}.
\newblock Technical report.
\newblock \urlprefix\url{http://alchemy.cs.washington.edu.}
\bibAnnoteFile{Kok09}

\bibitem{Selman93}
Selman B, Kautz H, Cohen B (1993) {Local search strategies for satisfiability
  testing}.
\newblock In: D J, M T, editors, Proceedings of Cliques, Coloring, and
  Satisfiability: Second DIMACS Implementation Challenge. American Mathematical
  Society.
\bibAnnoteFile{Selman93}

\bibitem{Wellman92}
Wellman MP, Breese JS, Goldman RP (1992) {From knowledge bases to decision
  models}.
\newblock Knowledge Engineering Review 7: 35--53.
\bibAnnoteFile{Wellman92}

\bibitem{Carter09}
Carter GW, Galas DJ, Galitski T (2009) {Maximal extraction of biological
  information from genetic interaction data}.
\newblock PLoS Computational Biology 5: e1000347.
\bibAnnoteFile{Carter09}

\end{thebibliography}

%\section*{Figure Legends}

%\begin{figure}[!ht]
%\begin{center}
%%\includegraphics[width=4in]{figure_name.2.eps}
%\end{center}
%\caption{
%{\bf Bold the first sentence.}  Rest of figure 2  caption.  Caption 
%should be left justified, as specified by the options to the caption 
%package.
%}
%\label{Figure_label}
%\end{figure}

%\section*{Tables}

%\begin{table}[!ht]
%\caption{
%\bf{Table title}}
%\begin{tabular}{|c|c|c|}
%table information
%\end{tabular}
%\begin{flushleft}Table caption
%\end{flushleft}
%\label{tab:label}
% \end{table}

\end{document}